\documentclass[journal]{IEEEtran}

\ifCLASSINFOpdf
\else
   \usepackage[dvips]{graphicx}
\fi
\usepackage{amsmath}
\usepackage{url}
\usepackage{graphicx}
\usepackage{soul}
\usepackage{xcolor}
\usepackage[ruled,vlined]{algorithm2e}
\usepackage{caption}
\usepackage{multirow}

\newcommand{\etal}{\textit{et al}.\@ }

\begin{document}

\title{Long Short-Term Memory Neuron Equalizer}
\author{Zihao Wang, \IEEEmembership{Member, IEEE}, Zhifei Xu, \IEEEmembership{Member, IEEE}, Jiayi He, Chulsoon Hwang, Jun~Fan,~\IEEEmembership{Fellow,~IEEE,}, Hervé~Delingette

\thanks{Manuscript received xxx. This work has been partially supported by the French government, through the 3IA Côte d’Azur Investments in the Future project managed by the National Research Agency (ANR) with the reference number ANR-19-P3IA-0002. And in part by the US National Science Foundation under Grant No. IIP-1916535. \textit{Corresponding author: Zhifei Xu.}}

\thanks{Zihao~Wang and Hervé~Delingette are with the French National Institute for Research in Computer Science and Control (INRIA) and Universit\'{e} C\^{o}te d’Azur (UCA), and Hervé~Delingette is with French Interdisciplinary Institutes of Artificial Intelligence (3IA Côte d'Azur). 06902 Valbonne, France (e-mail: {zihao.wang, herve.delingette}@inria.fr). }%
  \thanks{Zhifei~Xu, Jiayi He, Chulsoon Hwang and Fan~Jun are with the Electromagnetic Compatibility Laboratory, Missouri University of Science and Technology, Rolla, MO 65401, USA (e-mail: zxfdc, hejiay, hwangc, jfan@mst.edu).}
}
\markboth{Journal of \LaTeX\ Class Files, Vol. xx, No. 8, Oct 2020}
{Shell \MakeLowercase{\textit{et al.}}: Bare Demo of IEEEtran.cls for IEEE Journals}
\maketitle

\begin{abstract}
In this work we propose a neuromorphic hardware based signal equalizer by based on the deep learning implementation. The proposed neural equalizer is plasticity trainable equalizer which is different from traditional model designed based DFE. A trainable Long Short-Term memory neural network based DFE architecture is proposed for signal recovering and digital implementation is evaluated through FPGA implementation. Constructing with modelling based equalization methods, the proposed approach is compatible to multiple frequency signal equalization instead of single type signal equalization. We shows quantitatively that the neuronmorphic equalizer which is amenable both analog and digital implementation outperforms in different metrics in comparison with benchmarks approaches. The proposed method is adaptable both for general neuromorphic computing or ASIC instruments.
\end{abstract}


\IEEEpeerreviewmaketitle

\section{Introduction}
\label{sec:introduction}
\IEEEPARstart{N}{owadays}, the breakthrough of the massive data processing data factories leads the demand for data transmission efficiency from both macro-level (telecommunication),  meso-level (onboard systems), micro-level (inside chips). With the transmission speed increases dramatically, it is challenging to maintain the signal integrity (SI)[], power integrity (PI) in high-speed links design. The impedance discontinuities, field interaction, simultaneous switching noise (SSN), power delivery networks (PDN) and etc, will distort the high-speed signals, and therefore the intersymbol interference (ISI) which is a common phenomenon for high-speed data transmission systems will increase, the data output from the high-speed channel will hardly be recognized by the receiver. Thus, the equalization techniques have been developed to reconstruct the signal at the output of the high-speed channel.

\begin{figure}[hbt!]
    \centering
    \includegraphics[scale=0.41]{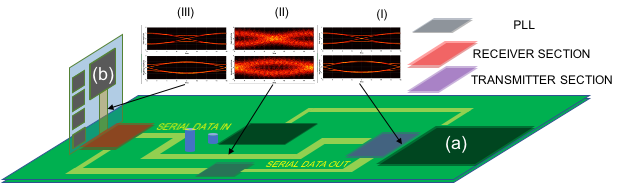}
    \caption{Backboard SerDes architecture. Signal emission from chips are transferred through a transmission line which leads to loss and distortion.}
    \label{fig:serdes}
\end{figure}

For a typical Serializer-Deserializer (SerDes) system targeted for high-speed data transmission as shown in Fig.~\ref{fig:serdes} which consistent by three modules: transmitter TX, receiver RX and phase locked loop (PLL) block. The signal is transmitted from TX through the channel and received by the receiver. The PLL is applied for clock signal synchronization which is a naive idea for identifying the delay of transmission channel to proper match the delayed signal.
In the RX block, the equalization techniques will be implemented to reconstruct the deformed signal due to the channel loss and other complex environmental influence. There are three types of equalizers are generally employed in the RX module. The feed-forward equalizer (FFE), the decision feedback equalizer (DFE) and the continuous time linear equalizer (CTLE). The FFE is dedicated to compensate the loss of the channel which is assumed to be a FIR filter with a transfer function inverse to the channel transfer function . Since only the compensation of the loss is not enough due the noises come from the complex environment. Therefore, the FFE and DFE are employed together to have better performance on the signal reconstruction. Since all the bit signals are the pulses, when the bit signal go through the channel, there will be delay and tails. The DFE is applied to drop the tail down to 0, then the pulse tail will not affect the next bit signal, thus, the better signal reconstruction can be reached. 
However, due to the design of channel and inherent non-linear representation limitations, the FFE and DFE combinations are unable to equalize original signals especially with complex channel in very high-speed data transmission. Facing to the new challenge, with the emergence of artificial intelligence technology \cite{AIinSignal}, the learning based equalizers are proposed.
\subsection{Learning based equalizer}
Modelling based equalizers are usual need a good understanding of the transmission systems attributions and are usual need rich expert knowledge for design and lack of generalization ability \cite{AdaNN}. Yet, the learning based approaches are more adaptable over varied channel condition. Most of the learning-based equalizers are no-linear equalizer. Some representative works \cite{ChenBayesian}\cite{ChernBayesianDFE}\cite{Kiguradze} were using a Bayesian statistical model for channel equalization. Neural network based approaches are used by some prior works \cite{oatao13123} \cite{RajbhandariNN} \cite{ZhangNN} \cite{AdaNN}for channels equalization. Some uncommonly used low speed equalization technologies such as Support Vector Machine based equalizer~\cite{TaubSVM} \cite{BiSVM}, Fuzzy based networks equalizer \cite{MayerFuzzy} were mostly difficult for hardware implementation  and are not elaborated in detail here.

\paragraph{Statistical learning}  Early work of Chen. \etal \cite{ChenBayesian} presented an approach for treating the equalization problem as the maximum \textit{a posteriori} with decision equalize. The Bayesian learning-based equalizer is outperformed than the LSE approach in quantitative performance evaluation. More rigorous, the Bayesian-based equalization process can be present with:
\begin{equation}
    P_{\theta}(s'_k|r_{k...k-n}) = \frac{P(s'_k, r_{k...k-n})}{P(r_{k...k-n})}
    \label{eq:bayesian}
\end{equation}
where $r_{k...k-n+1}$ is a vector of last n observation signals of the receiver side and $s_k$ is the equalized signal of time $k$. The Eq.~\ref{eq:bayesian} shows the symbols equalization problem of $s_k$ can be modelled as the probability inference of symbols. The symbols discriminated based on the \textit{a posterior} probability $P(s'_k|r_{k...k-n})$ which can be computed from the \textit{a prior} probability $P_{\theta}(r_{k...k-n})$ and the joint probability of the observation from transmission channel $P(s'_k, r_{k...k-n})$. This Bayesian model can be trained by optimizing the probabilities models parameters $\theta$ through maximizing the log-likelihood function with gradient-based optimization methods. 

\paragraph{Neural network learning}
Recent work of Chu \etal \cite{ChuNN} introduced a neural network circuitry equalizer (NNE) for high-speed signal link construction. An artificial neural network (ANN) is integrated into the continuous-time linear EQ (CTLE) as the replacement of the DFE block (block (b) in Fig.~\ref{fig:serdes}). Their quantitative experimental shown that the ANN-based equalizer is more robust than the DFE approach for degraded signal restoring accuracy. In terms of the transmission speed improvement, the NNE improved about 30\% in data rate vis-à-vis the traditional equalization method. A similar work of Rajbhandari \etal employed ANN for multiple-input-multiple-output visible light communication (MIMO-VLC) system signal equalization. The MIMO-VLC equalization problem is usually more challenge than the PCB level signal equalization as the channel of MIMO-VLC system can be an open environment which may be easily violent by the external optical noisy. The experimental study showed that the ANN equalizer outperformed than the temporal-spatial cross-talks compensation\cite{RajbhandariNN}.

Many prior works have shown that the potentiality of neural network based equalization for the future high-speed, complexity electromagnetic channels and long distance transmission systems design. We see that most of the neuromorphic computing systems used for equalization are still limited to the basic perceptions or multi-layer perceptron neural systems. 

\subsection{LSTM Equalizer and Deep Learning Approach}
In this paper, we propose a novel approach based on the long short-term memory neural network and deep learning for channel signal equalization. In many circumstances, neural networks computing are software implemented and run on GPGPU (General-purpose processing on graphics processing units) or TPU (Tensor Processing Unit) acceleration which limited the application of neural networks for highly integrated circuit systems. Yet, the proposed approach can be implemented through programmable gate array or ASIC or general neuromorphic computing instruments (related literature: \cite{SPMneuronmorphic, YannNeuromorphic, BalajiNeuromorphic}) which realize the embedding for integrated system. To the best of our knowledge, our proposed method is the first application of physical realizable temporal memory based neural equalizer. 

Bayesian DFE achieves a reasonable performance in terms of Bit Error Rate (BER) in comparison with the conventional DFE\cite{ChenBayesian}. However, the Bayesian DFE is not amenable for variable frequency channel signal equalization as the equalization model is constrained to the optimization of a unique frequency and the relative weaker nonlinear transformation ability. Yet, the proposed approach is adaptable to variable frequency signal equalization with the powerful approximation ability of neural networks.

The conventional ANN based equalizer is not capable for capturing the time sequential information as the input signals are processed with out considering the temporal relationship.\cite{JingDai} However, the signal equalization task is a typical sequential processing problem which is suitable for employing our LSTM based equalizer. 

The paper is organized as follows: In section \ref{sec:lstme} we first describe the memory gates mechanism of LSTM network and the construction procedures for LSTM based equalizer architecture design. Section \ref{sec:experiment} describes the implementation of the proposed method both simulated and physical realization. The conclusion section \ref{sec:conclusion} summaries the full paper and look into the future of applying deep learning approaches for equalizer design.


\begin{figure}
    \centering
    \includegraphics[scale=0.25]{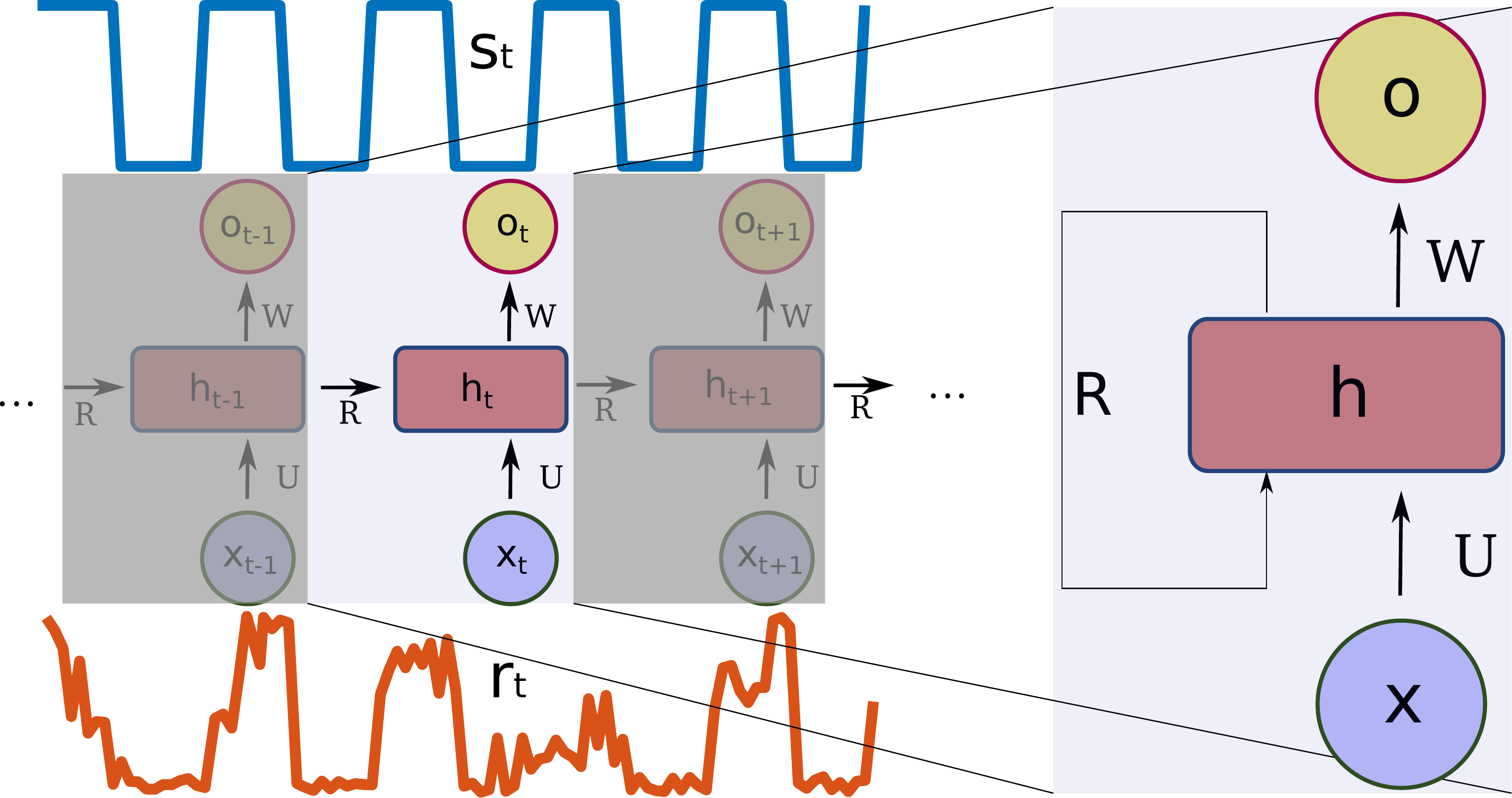}
    \caption{Unrolling demonstration of a Recurrent neural network for time series processing. Input signal is red series $r_t$ and output signal is shown with blue line $s_t$}
    \label{fig:rnn}
\end{figure}
\section{Long short-term memory neural equalizer}
\label{sec:lstme}
\subsection{Long Short-Term memory neural network}
\begin{figure}
    \centering
    \includegraphics[scale=0.2]{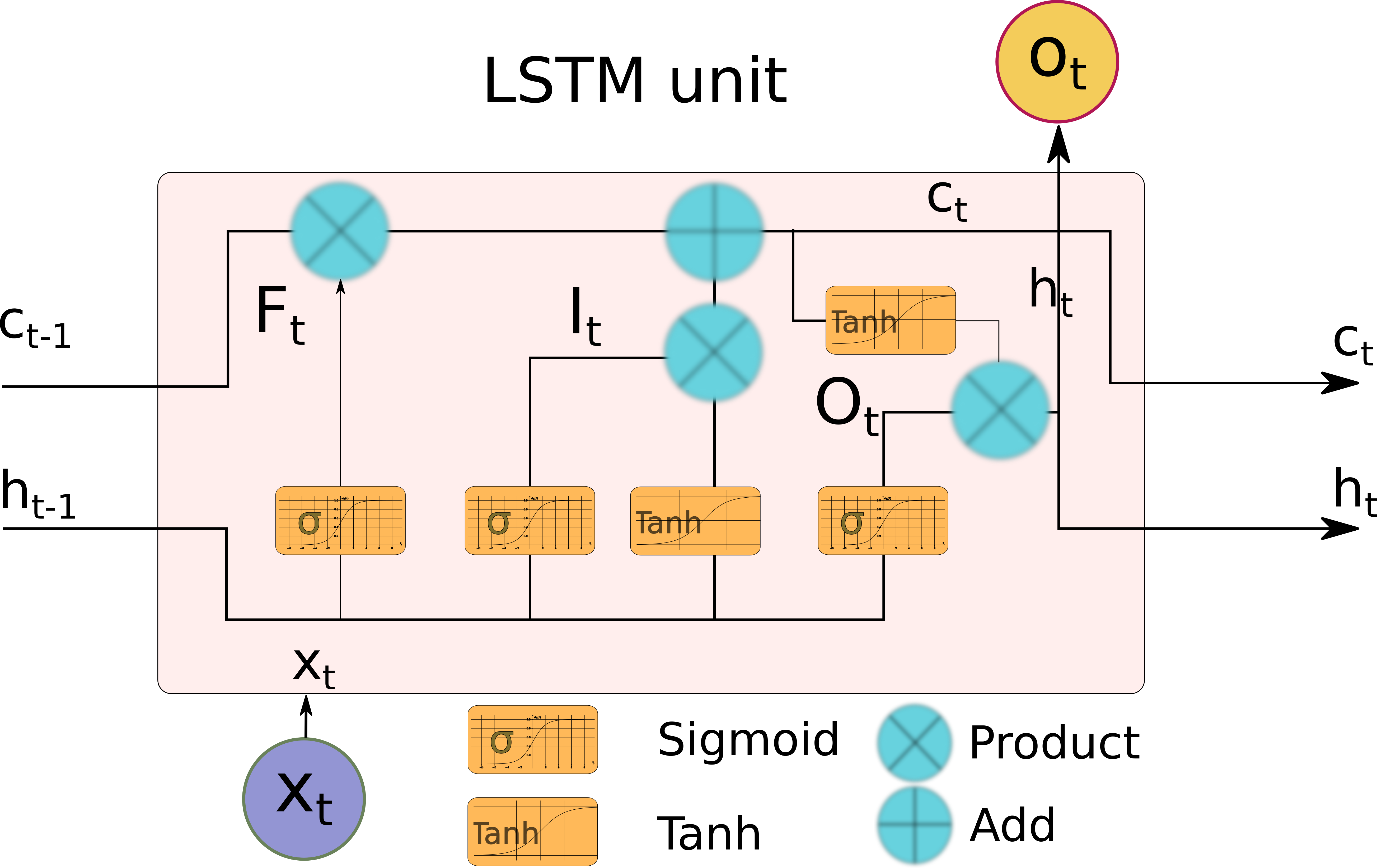}
    \caption{Architecture of a LSTM cell. The cell's long memory is $c_{t-1}$ and hidden state is $h_{t-1}$. At time $t$, the input signal of current state $x_t$ is processed by the LSTM cell to generate the current hidden and cell state.}
    \label{fig:lstm}
\end{figure}
The long Short-Term memory neural network (LSTM)\cite{SchmidhuberLSTM} is a type of recurrent neural networks (RNN) which is specially designed for processing the time series or sequential data. Differ from the full connected neural network, the RNN can learn the temporal information by recurrent backpropagation\cite{RumelhartRNN}. Fig.~\ref{fig:rnn} shows a RNN for signal series regularization. At time $t$ the RNN is feed with the noisy signal $x_t = r_t$ and the memory information $R$ passed from the last time $t-1$, then the neural $h_t$ will generate the processed signal $O_t = s_t$ of time $t$. Yet, RNN is limited by the memory information transmission flaw that the memory passed from the last time $t-1$ might be not sufficient for processing the input of current time $t$. The LSTM is proposed to tackle this problem by introducing the long short-term memory gates mechanism\cite{SchmidhuberLSTM}.
It has been practically proved that the LSTM is better than RNN for long-distance memory and the successful application of LSTM in many areas related to time series signals such as Google neural machine translation system, Microsoft voice recognition system, ECG beats signal classification and Apple's Siri assistant systems etc\cite{googleLSTM}\cite{li2015lstm}\cite{ECG}. 
\begin{figure*}[ht!]
    \centering
    \includegraphics[scale=0.3]{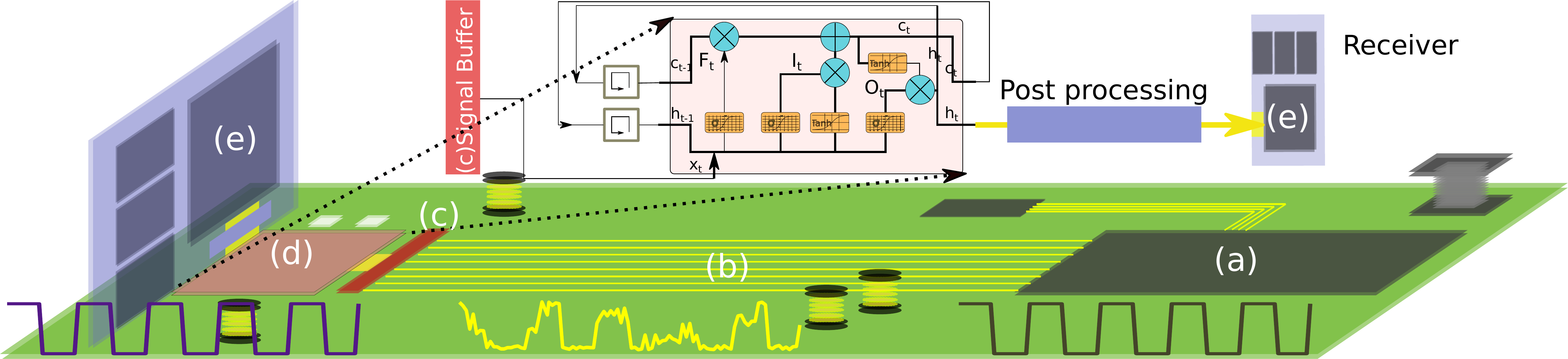}
    \caption{The signal link of the proposed LSTM neural network based channel equalizer. (a) Chip generates logic bits. (b) Arbitrary signal channel. (c) N bits signal delay. (d) LSTM neural network signal interpreter. (e) Signal receiver.}
    \label{fig:lstmEqualizer}
\end{figure*}

As Fig.~\ref{fig:lstm} shows, the LSTM cell consists a group of neurons and several information transformation flows where $c_{t-1}$ is cell state variable and $h_{t-1}$ indicates the hidden state variables at time $t-1$. Cell state variable carrying the memory information of the previous cell's state to the current state and the hidden variables are the cell output information of the current cell. Inside the LSTM cell, two activation functions are used for non-linear transformation of the information flows. The Sigmoid function:~$\sigma(x) = \frac{1}{1+exp{(-x)}}$ is used to compute $\sigma(h_{t-1})$ which controls the forget gate $F_t$ by a Hadamard product with the last cell state $c_{t-1}$ and the input gate $I_t$. The activation level of the forget gate is correlated with the information passing degree from the last cell state. The tanh function:~$tanh(x) = \frac{exp{(2x)} - 1}{exp{(2x)} + 1}$ is used for memory selection by controlling the activation level of the current cell input state $x_t$. The signal gated by the two different signal flows is then summed together as the current cell state $c_{t-1}$ and a Tanh non-linear transformation of the current cell state $c_t$ and sigmoid non-linear transform of the last cell hidden state $h_{t-1}$ are multiplied together with Hadamard product to generate current cell state signal $h_t$. In summary each of the gates in Fig:~\ref{fig:lstm} are given by:
\begin{equation}
\begin{aligned}
    f_t = \sigma(w_f x + w_{rf} h_{t-1} + b_f) \\
    i_t = \sigma(w_i x + w_{ri} h_{t-1} + b_i) \\
    cs_t = \tanh{(w_c x + w_{rc} h_{t-1} + b_c)} \\
    o_t = \sigma(w_o x + w_{ro} h_{t-1} + b_o) \\
\end{aligned}
\end{equation}
they are forget gate $f$, input gate $i$, cell candidate $cs$ and output gate $o$, respectively.
With the governing of the four information control gates, the output of the current cell states is given by,
\begin{equation}
    \begin{aligned}
        c_t = f_t \otimes c_{t-1} + i_t \otimes cs_t \\
        h_t = \sigma_t \otimes \sigma_c(c_t)
    \end{aligned}
\end{equation}
The neural networks weigh parameters $W=\{\{w^n_f,w^n_i,w^n_c,w^n_o]\}, n\in\mathcal{N}\}$ and the bias parameters $b=\{\{b^n_f,b^n_i,b^n_c,b^n_o]\}, n\in\mathcal{N}\}$  inside the LSTM cell are trainable variables. A recurrent weights $W_r=\{\{w^n_{rf},w^n_{ri},w^n_{rc},w^n_{ro}]\}, n\in\mathcal{N}\}$ is also trainable for LSTM networks. 
The LSTM can easily be implemented by digital circuit with adjustable neural activation parameters and the LSTM circuit is amenable for analog computing which the neurons are directly modelling by electronic signals~\cite{JHanLSTMDigital}\cite{ZhaoLSTMCiruit}\cite{SmagulovaSurvey}. The analog neural networks implementation can break the gap of frequency limitation of equalizers where the only barrier is the frequency of the converter of the digital and analog signal (DAC).

\begin{figure*}[hbt!]
    \centering
    \includegraphics[scale=0.2]{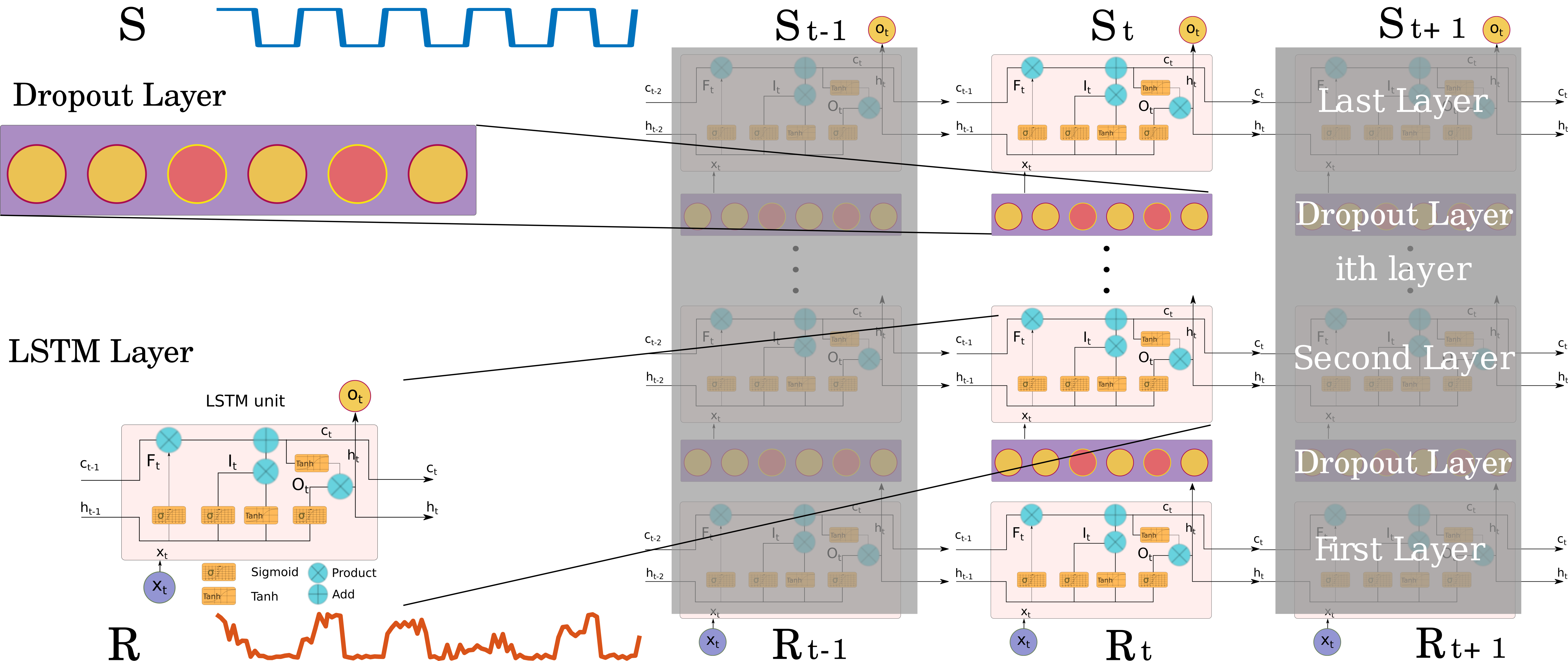}
    \caption{Deep Learning LSTM Equalizer consists many layers of LSTM neurons and dropout layers. The dropout layer are used for introducing the sparsity for the deep neural network.}
    \label{fig:DeepNeuron_DFE}
\end{figure*}
\subsection{Long short-term memory neural equalizer}
\begin{algorithm}[h!]
\SetAlgoLined
\SetKwInput{KwData}{Inputs}
\SetKwInput{KwResult}{Output}
\KwData{N bit Signal delayed: $r^n_{mem}$}
\KwResult{Equalized signal: $s'$}
Restore neural parameters form RAM: $W,WR,b$\tcp*{LSTM} 
Initialize states cache: $h=\mathbf{0}, c=\mathbf{0}$\tcp*{} 
\While{Clock Refresh}{
$t \gets clock~signal$\\
$h_{t-1}, c_{t-1} \gets h, c$ \tcp*{Get last cell states} 
$r_{t} \gets r^n_{mem}$ \tcp*{Get current signals} 
\tcp{Compute the four cell gates based on the input signal vector $r_t$ and last states $\{h_{t-1}, c_{t-1}$\}} 
$f_t \gets  \sigma(w_f r_t + w_{rf} h_{t-1} + b_f)$ \tcp*{forget gate} 
$i_t  \gets  \sigma(w_i r_t + w_{ri} h_{t-1} + b_i)$ \tcp*{input gate} 
$cs_t  \gets  \tanh{(w_c r_t + w_{rc} h_{t-1} + b_c)}$ \tcp*{cell gate} 
$o_t  \gets  \sigma(w_o r_t + w_{ro} h_{t-1} + b_o)$ \tcp*{output gate} 
\tcp{Compute the current cell outputs: $c_t, h_t$} 
$c_t \gets f_t \otimes c_{t-1} + i_t \otimes cs_t $\tcp*{cell gate} 
$h_t \gets \sigma_t \otimes \sigma_c(c_t)$\tcp*{hidden gate} 
$h, c \gets h_{t}, c_{t}$ \tcp*{Refresh states cache} 
$s' \gets \mathcal{F}(h)$\tcp*{Full connected layer}
$s' \gets \mathcal{FIR}(s')$\tcp*{Post processing filtering}
}
\caption{Signal forward propagation at time $t$.}
\label{alg:forward}
\end{algorithm}
The Long short-term memory neural equalizer is a intrinsic-feedback equalizer system where the spoiled signal series $r$ from the transmission line are directly processed by the LSTM neural block while the feedback of information is achieved by the long-shot memories states $h, c$. As Fig.~ \ref{fig:lstmEqualizer} shows the LSTM is integrated as a terminal block in the signal channel link, where the input sequential signals $r$ comes from the transmission channels (b) are resampled by a sample and hold circuit and stored in a $n$ bits memory buffer $r^n_{mem}$ (see block (c) of Fig.~\ref{fig:lstmEqualizer}). This buffer is a Serial-in parallel-out (SIPO) buffer where the first register $r^1_{mem}$ stores the $r_t$ and the last register stores the $r_{t-n}$.
The signals stored in the buffer are input to the LSTM interpreter (d) to generate the equalized signal. The equalized signals are post-processing by a signal processing block to get the equalized bits $s'$. A detailed presentation of the forward propagation of the signals at the time $t$ is demonstrated with algorithm \ref{alg:forward}.

The output hidden state $h_t$ at time $t$ is decoded by a full connected neuron $\mathcal{F}(h_t)$ to get the equalized signal output. This full connected cell is parameterized by $w_{full}$ and $b_{full}$. The output of the full connected cell is given by:$\mathcal{F}(h_t) = \sigma(w_{full} \dot h_t + b_{full}$. At the end, the signal is processed by a FIR filter to remove the noise and stabilizing the waveform.
The LSTM network aims to minimize the difference between the equalized bits $s'$ and the bits signal $s$ emitted from the chip (a). We can train the LSTM by minimizing the Mean-Square-Error (MSE) of the two items:
\begin{equation}
    MSE = \sum^N_{i-1}\frac{(s_i - s'_i)^2}{N}
    \label{eq:mse}
\end{equation}
The LSTM network parameters can be trained through backpropagation algorithm\cite{SCHMIDHUBER201585} to find the corresponding neural network parameters.
\subsection{LSTM training}
\paragraph{Dataset} A training dataset and validation dataset need to be constructed for training the LSTM equalizer. The neural network can be trained though a batch of training input $B_t$ and corresponding target signal. The validation dataset is used to avoid overfitting problem of the learning based methods. We denote the training set as $\mathcal{D}^K_{train}:\{S_{train}|R_{train}\}$,
\begin{equation}
\centering
R_{train}=
   \begin{Bmatrix}
    \{R_{t-n} &...& R_t\}^1\\
    \{R_{t-n-1} &...& R_{t-1}^2\}\\
    &...& \\
    \{R_{0} &...& R_n\}^K\\
    \end{Bmatrix} ;
S_{train}=
    \begin{Bmatrix}
    \{S_t\}^1\\
    \{S_{t-1}^2\}\\
    ...\\
    \{S_n\}^{K} \\
    \end{Bmatrix} 
\end{equation}
where the $R_{t-n} ... R_t$ represent a sequence of slice of the sequential signals which is terminated at time $t$. and same notation for the validation dataset $\mathcal{D}^K_{valid}:\{S_{valid}|R_{valid}\}$.
The dataset can be measured from the physical transmission platforms or be collected from a simulation of the transmission channels in case the s-parameters of the transmission line is known. In the training phase, the $\mathcal{D}^{K'}_{train}$ is feed to the LSTM network for minimizing the MSE loss function by stochastic gradients descent based algorithms (\textit{i.e.:} Adam in this work)\cite{Kingma2015AdamAM}.
\begin{algorithm}[h!]
\SetAlgoLined
\SetKwInput{KwData}{Inputs}
\SetKwInput{KwResult}{Output}
\KwData{Training data : $\mathcal{D}^K_{train}:\{S_{train}|R_{train}\}$}
\KwData{Validation data : $\mathcal{D}^{K'}_{valid}:\{S_{valid}|R_{valid}\}$}
\KwData{Learning rate:$\alpha$; Validation Steps: $V$}
\KwResult{LSTM Parameters: $W,WR,b$}
\tcp{Initialize with Xavier method\cite{xavier}} 
Xavier initializer: $W,WR,b$\tcp*{} 
Initialize states: $h=\mathbf{0}, c=\mathbf{0}$\tcp*{} 
\While{$\epsilon_{valid}<\delta_{converge}$}{
    \For{$i = 1;i < K;i++$}{
    $R^i \gets \{R_{t-n} ... R_t\}^i$ \tcp*{Get ith input}
    $S^i \gets \{S_t\}^i$ \tcp*{Get ith training target}
    \tcp{Forward propagation with Algo:~\ref{alg:forward}}
    $R' \gets LSTM(R^i)$\tcp*{}
    \tcp{Get the loss between output and target based on loss function:\ref{eq:mse}}
    $\epsilon = loss(R', S^i)$\tcp*{}
    $\Delta_{W} \gets \alpha \times \frac{\partial \epsilon}{\partial W}$\tcp*{Jacobian W}
    $\Delta_{WR} \gets \alpha \times \frac{\partial \epsilon}{\partial WR}$\tcp*{Jacobian WR}
    $\Delta_{b} \gets \alpha \times \frac{\partial \epsilon}{\partial b}$\tcp*{Jacobian b}
    $W \gets \Delta_{W} + W$\tcp*{Update W}
    $WR \gets \Delta_{WR} + W$\tcp*{Update WR}
    $b \gets \Delta_{b} + W$\tcp*{Update b}
    \eIf{$0 \equiv i \mod V$}{
        $\epsilon_{valid} \gets 0$ \tcp*{}
        \For{$j = 1;j < K';j++$}{
        \tcp{Get validation dataset}
        $R^j \gets \mathcal{D}^{K'}_{valid}: \{R_{t-n} ... R_t\}^j$ \tcp*{}
        $S^j \gets \mathcal{D}^{K'}_{valid}: \{S_t\}^j$ \tcp*{}
        \tcp{Forward propagation with Algo:~\ref{alg:forward}}
        $R' \gets LSTM(R^j)$\tcp*{}
        $\epsilon_{valid} \gets \epsilon_{valid} + loss(R', S^j)$\tcp*{}
            }
        $\epsilon_{valid} = \frac{\epsilon_{valid}}{K'}$\tcp*{}
        $break$\tcp*{}
        }
        {$continue;$}
    }
 }
\caption{LSTM training based on backpropagation.}
\label{alg:feedback}
\end{algorithm}
\paragraph{Backpropagation} The parameters of the LSTM equalizer are determined through a batch based backpropagation algorithm \cite{SCHMIDHUBER201585} which can be interpreted as two phases: (I) foreword propagation as we shown in Alg:~\ref{alg:forward}. (II) backward propagation where the neural networks parameters are updated based on the gradients information of errors feedback. The details of backpropagation are described in Alg:~\ref{alg:feedback}
\subsection{Variable-frequency Signal Equalization}
The proposed LSTM equalizer is amenable for processing variable-frequency signals. Yet, as the complexity of the input signal is much higher than a single frequency signal, the single layer LSTM equalizer maybe not capable to learn the variant conditions or might need along time to achieve a usable accuracy. Deep learning is a effective approach for improving the abstraction and memory ability of the LSTM neural netowk. In another aspect, the LSTM have generalization ability to process the certain range of frequency variation. Yet, as the range of the input signal frequency is fair large (\textit{i.e. 10 GHz$\sim$60 GHz}), the LSTM equalizer is no longer capable to correct the signal. As the variable-frequency signals are more complex than the fix frequency signal. The LSTM equalizer should be improved for adopting more complex signal processing environment. Theoretically the input signal delay do not need modify as the LSTM neural network can memory both the long and short information from the past signals the bias comes from the frequency variance can be compensated by the LSTM. However, in case the frequency change range bias far away from the designed work frequency, the LSTM will no longer capable for the equalization. In this sense, the input signal delay of LSTM equalizer need to be adjusted in order to cover a sufficient memory range. There are two ways to adapt the wide range frequency equalization for LSTM equalizer: (I) increasing the resolution of delay points for high frequency signal which need a modification of the hardware wiring. (II) add a signal preprocessing block before transmit the signal to LSTM equalizer. To coup with the two aspects, we first introducing the deep learning for LSTM equalzier and then discussing the preprocessing block design in the after.
\subsubsection{Deep Learning in LSTME}
\begin{figure*}[ht!]
    \centering
    \captionsetup{justification=centering}
    \includegraphics[scale=0.67]{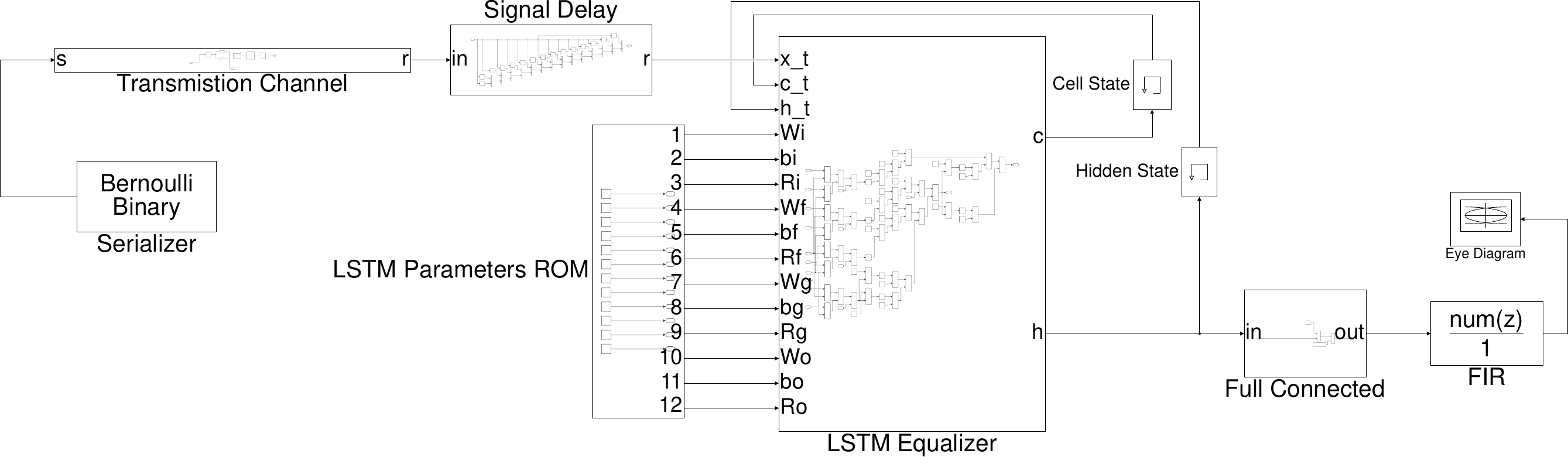}
    \caption{The Simulink architecture diagram of the proposed LSTME applied for a simulated channel signal equalization.}
    \label{fig:deepLSTM}
\end{figure*}

\begin{figure}[h]
    \centering
    \includegraphics[scale=0.65]{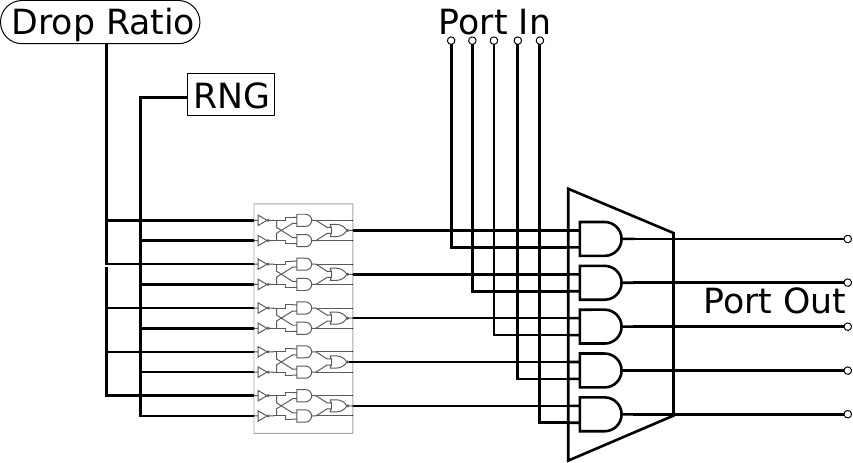}
    \caption{Dropout layer for a single digital feature channel (encoder circuits are omitted). Parallel features channels can be realized by the combination of many single circuits.}
    \label{fig:dropout}
\end{figure}

Deep learning is an effective mythology for constructing the neural networks with a deep stack structure in order to get a more powerful non-linear representation ability for traditional neural systems\cite{lecun2015deeplearning}\cite{SCHMIDHUBER201585}. The LSTM equalizer can also be improved through combining several LSTM equalizers for further improving the equalization ability. Theoretically, a deeper LSTM equalizer system can accommodate the more worse situation that the channel output signal are strongly non-linear and heavily spoiled by external noise. Fig.~\ref{fig:deepLSTM} shows a skeleton structure of a deep learning LSTM stack architecture. A deep lstm equalizer consists several layers of LSTM networks with dropout layers~\cite{dropout} following. The dropout layers are applied for avoiding the overfitting (regulator) problem of the deep neural network training. As the Fig.~\ref{fig:dropout} shows a digital implementation of single feature channel of a dropout neural layer.\cite{dropoutHardware} The random number generator (RNG) generates a random number and the comparators compare the generate number with the drop out ratio to determinate the feature channel is enable or not.
The enabled feature channels are then passing the features from previous layers to next LSTM layer. As the number of layers increasing the ability of equalization of the deep LSTM equalizer will be improved. 
\subsubsection{Deep LSTME towards Variable-frequency}
The deeper LSTME is more capable for high order non-linear signal processing as the feature extraction and reasoning ability of the LSTM is improved through the deep learning design. To make the deep LSTM equalizer amenable for variable frequency signal equalization, the deep LSTM need to be train on the sampling dataset that covering the frequency variation range. The dataset contains the channel output signal and input signal across several selected frequency in the range.
The advange of using deep LSTM equalizer for variable frequency signal equalization is no hardware modification is required for adapting the multiples frequency and the equalizer can adapt difference frequency as the same time. Yet, an adjustment of the parameters in FFE-DFE system is necessary to do this and the re-tuned parameters of the FFE-DFE equalizer is no longer effective on the previous frequency.

\subsection{LSTME implementation}
The proposed LSTM equalizer system is implemented and verified through Simulink modelling. As shown in Fig.~\ref{fig:deepLSTM}, the signals are generated by a Bernoulli binary generator. Signals are transmitted through a transmission channel and then delayed by the signal delay circuits as the input to the LSTM equalizer. The trained LSTM neural networks are stored in the LSTM parameters ROM and the LSTM equalizer will read those parameters to determinate the output signals: cell states signal and hidden states signal. Both of the two signal flows are feedback to the LSTM as inputs for next signal equalization. The cell state signal is further processed  by the full connected neuron block to generate the response signal of the LSTME. in the final stage the FIR filter is applied for the final signal correction.
\section{Experiment and Result}
\label{sec:experiment}
Two experiments have been conducted and compared with optimized FFE-DFE. The first case is mainly focused on the waveform of the channel output at 10 Gbps speed, a simulation based channel is employed. The second case is focused on the speed. A designed PCB channel characterized up to 50GHz is employed. The data rate is simulated up to 50Gbps. For both cases, the waveform and the characteristics of eye diagrams are compared. 
\begin{figure}[h!]
    \centering
    \includegraphics[scale=0.45]{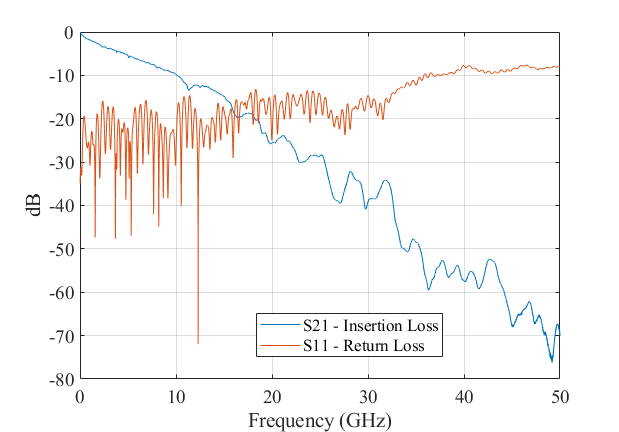}
    \caption{Return loss and insertion loss of the simulated channel.}
    \label{fig:S_para}
\end{figure}
\subsection{Experimental result for a high-speed PCB channel}
A fabricated PCB channel, containing transmission line, via connector, etc, has been used for this experiment. Due to non-disclose agreement, the parameters of the channel are not shown in this paper which are not important data for our proposed method. However, the S parameter of the channel is shown and characterized up to 50 GHz. The insertion loss and return loss of the S parameter are shown in Fig. \ref{fig:S_para}. The insertion loss exposes very high loss in high frequencies. The transmitted signal will be attenuated a lot and therefore hard to be recognized by the receiver. Therefore, the FFE-DFE or our proposed method is need to recover the signal.

\begin{figure}[h!]
    \centering
    \includegraphics[scale=0.45]{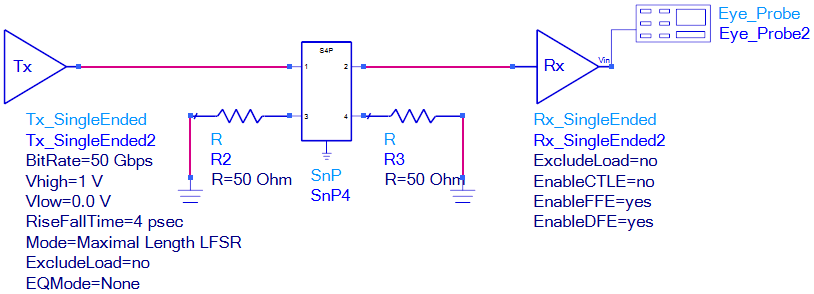}
    \caption{Schematic of the ADS channel simulation.}
    \label{fig:ADS_schematic}
\end{figure}

\begin{figure}[h!]
    \centering
    \includegraphics[scale=0.465]{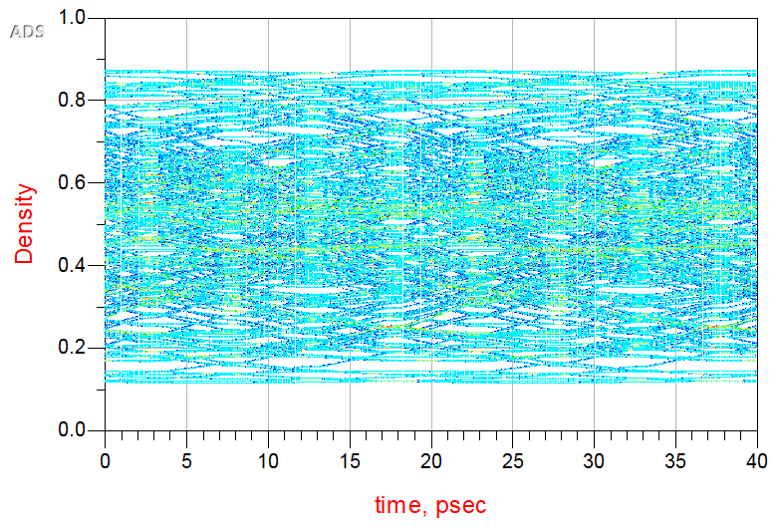}
    \caption{Eye diagram of ADS channel signal simulation (raw signal from channels).}
    \label{fig:no_dfe}
\end{figure}

\begin{figure}[h!]
    \centering
    \includegraphics[scale=0.465]{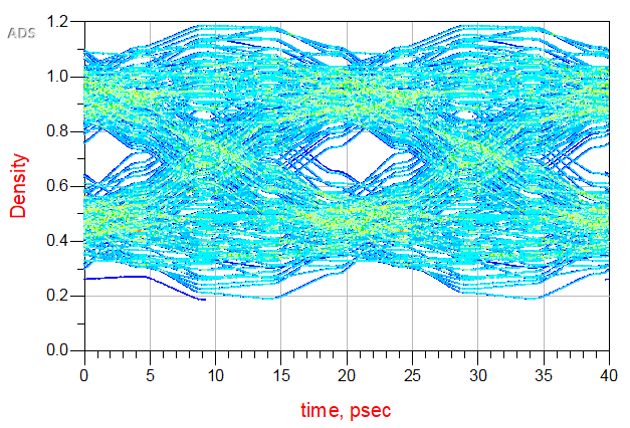}
    \caption{Eye diagram of ADS channel signal equalized via the conventional active FFE-DFE system.}
    \label{fig:ffe-dfe}
\end{figure}

A simulation is performed in ADS with the S parameters of the fabricated PCB channel. The schematic is shown in Fig. \ref{fig:ADS_schematic}. Since there are two channels on the board, only one channel is employed in this simulation. A 50 Gbps data rate is chosen for the simulation. The high/low levels are defined as 1 $V$/0 $V$, respectively. The rise and fall times are equal to 4 $psec$. 
The simulated channel without FFE-DFE is shown in Fig. \ref{fig:no_dfe}. Fig. \ref{fig:ffe-dfe} shows the output results with optimized FFE and DFE. The FFE and DFE cursors are shown in Table. \ref{tab:DFE}.
\begin{table}[]
\centering
\caption{DFE and FFE parameters}
\label{tab:DFE}
\begin{tabular}{cc|c}
\hline \hline
\multicolumn{2}{c|}{\textbf{FFE}} & \textbf{DFE} \\ \hline
PreCursor & PostCursor & Taps \\
-2.337340 &  4.038660 & 0.322812 \\
 &  & -0.017401 \\
 & -2.185680 & 0.048581 \\
 0.782150 &  0.534350 & -0.065590 \\
 &  & 0.039204 \\
 & -0.121820 & -0.021085 \\ \hline \hline
\end{tabular}
\end{table}

As shown in Fig. \ref{fig:no_dfe}, the output data is hardly to be recognized by the receiver without the FFE-DFE equalization. Yet, with the FFE-DFE equalizers processing, the eye is opened slightly with a fair large jitter as shown in Fig. \ref{fig:ffe-dfe}.

\begin{figure}[h!]
    \centering
    \includegraphics[scale=0.465]{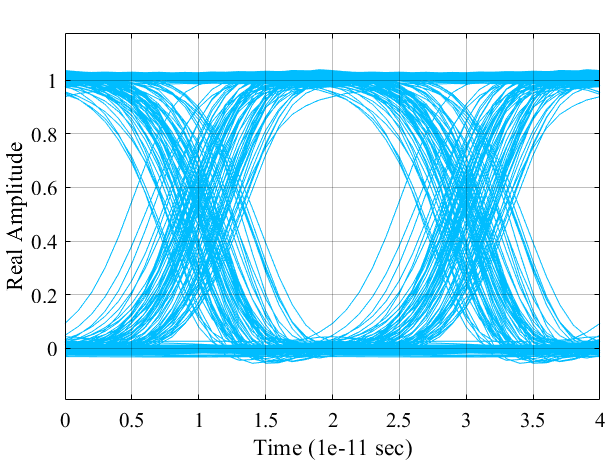}
    \caption{Eye diagrams of ADS channel signal (50 GHz) equalized via the proposed LSTM equalizer system.}
    \label{fig:EYE50GHZDFE}
\end{figure}

Fig.~\ref{fig:EYE50GHZDFE} shows the eye diagram of the equalized result of the LSTM equalizer. We see that both the eye width and heights are outperformed than the FFE-DFE approach and the signal jitter is significantly reduced.

\begin{figure}[h!]
    \centering
    \includegraphics[scale=0.49]{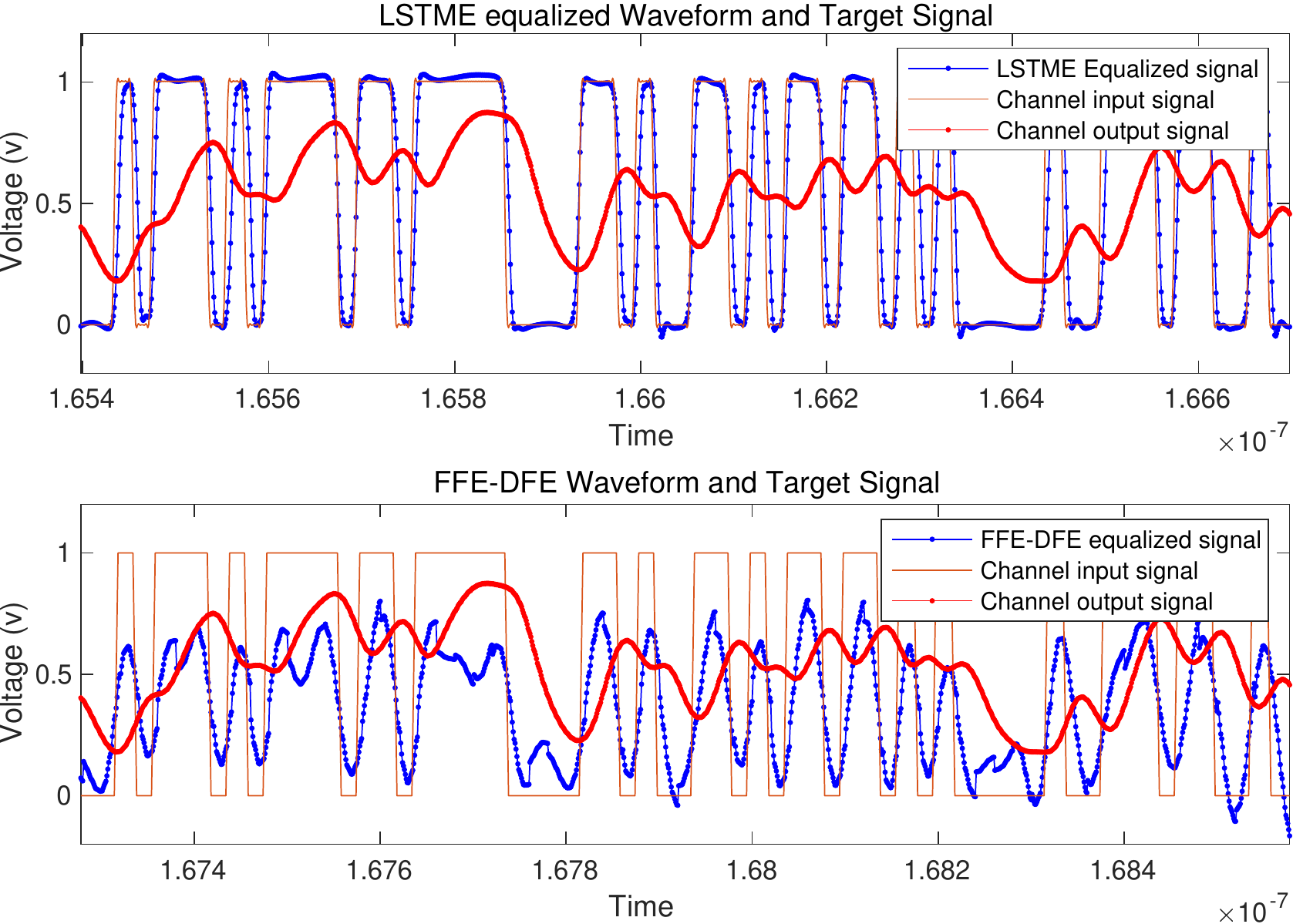}
    \caption{Comparison between LSTME and FFE-DFE equalized signal waveform with input and output signal of the transmission channel shown.}
    \label{fig:waveform}
\end{figure}

The wave form comparison is shown in Fig. \ref{fig:waveform}. The red curve is the output signal without any equalization. The blue signal in the lower part figure is the output signal after active FFE/DEF equalization. And the blue signal in the upper figure is the output signal after our proposed equalizer. The results clearly shows that our proposed method has a very good signal recover ability. 

\subsection{Experimental results for a mismatched channel}
In the world of signal integrity, due to the limitation of PCB manufacturing, the impedance mismatching is a general problem for the high speed transmission system designs. The overshoot and undershoot problems will appear due to the impedance mismatching. And the defect on the channel will also distort the output signals. Therefore, in this section, our proposed equalizer is proposed to a bad designed channel with mismatched loads to present the strength of our method. The eye diagram is first obtained through ADS simulation with optimized FFE and DFE settings. A 20 Gbps signal with 10 ps rise/fall time is applied to this bad designed channel. 
\begin{figure}[h!]
    \centering
    \includegraphics[scale=0.4]{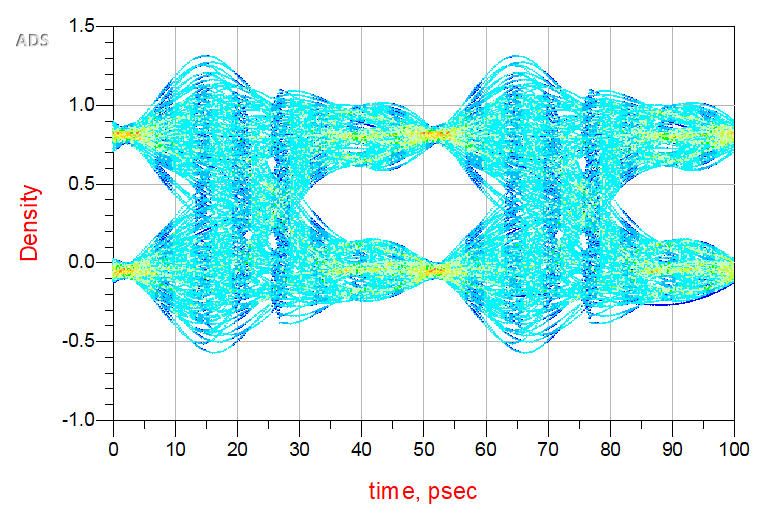}
    \caption{ADS simulated eye diagram with bad designed channel}
    \label{fig:res-dfe}
\end{figure}

\begin{figure}[h!]
    \centering
    \includegraphics[scale=0.45]{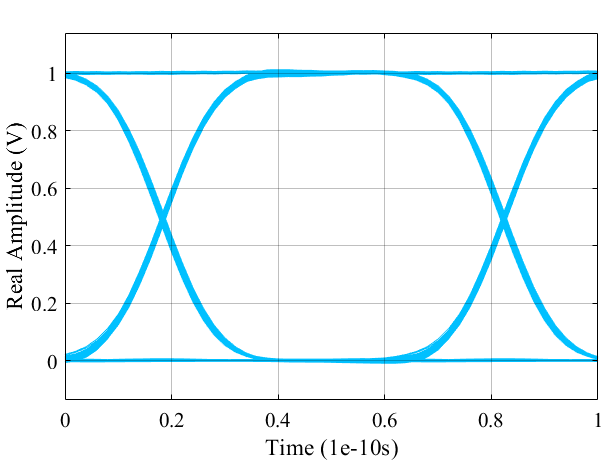}
    \caption{Eye diagrams of bad designed channel signal (20 GHz) equalized via the proposed LSTM equalizer system.}
    \label{fig:eye20Ghz}
\end{figure}

\begin{figure}[h!]
    \centering
    \includegraphics[scale=0.5]{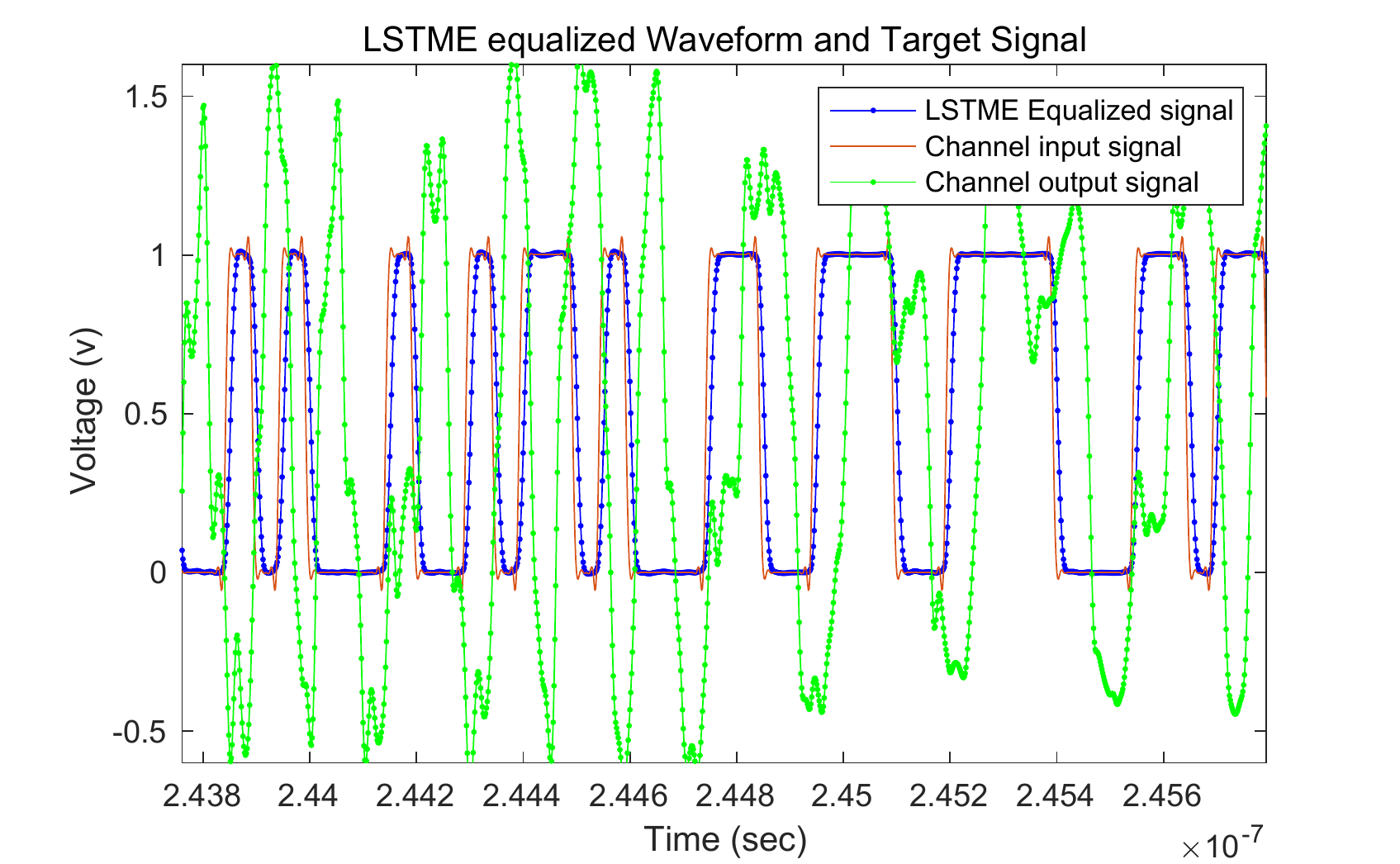}
    \caption{Waveform of LSTM equalizer for bad designed channel of 20GHz signal.}
    \label{fig:20GHZwave}
\end{figure}
\section{Discussion}
\label{sec:discussion}
The proposed LSTM equalizer is a neural learning based signal equalizer without necessary to do signal pre-processing (FFE block). Current widely used solution for the transmission line signal recovering are general realized though DFE or FFE-DFE combination\cite{WangFFEDFE,AhnFFEDFE, ZhengFFEDFE}. The LSTM addressed to a novel learning based equalizer which not only more suitable for high non-linear signal restore but adaptable for variable-frequency signal equalization for the transmission line.
The effectiveness of the LSTM equalizer have been shown through a ADS simulation channel signal equalization task with quantitative and qualitative comparison with a FFE-DFE combination.

The LSTM neural network shows reasonable equalization result in comparison with the FFE-DFE combination. Facing to a complexity sense of variable frequency signal equalization. A deep learning based LSTM equalizer and corresponding implementation shows the effective of the deep LSTM equalizer for variable-frequency signal equalization. The advantage of trainable LSTM equalizer is the learning parameters are customizable which can be applied to flexible and more complex scenario without hardware modification. This can reduce the equalizer implant cost for variant transmission channels and introduce portability for practical application.

The use of LSTM equalizer result in better regularized output waveform as we shown Fig:\ref{fig:waveform}. Nevertheless, for FFE-DFE based equalization approaches the restored transmission signals are still spoiled and very different from the original signal. The better regularized signal of LSTM equalizer can reduce the design complexity of afterward signal decision module. 

A limitation of LSTM equalizer lies in the relative computational complexity in comparison with traditional FFE-DFE combination. This flaw is due to the excessive requirement of non-linear representation capability of the deep neural networks. Empirically, we noted that in section \ref{sec:experiment} for the equalization of a fabricated PCB channel with a fix frequency 50 $Gbps$ input signal, a LSTM equalizer with 20 hidden cells and 15 delays with resolution of 5 $ps$ is enough for getting an out-performed equalized output. In this scenario the computational complexity is acceptable in practical usage. However, the complexity of a deep learning LSTM equalizer can be fair high as the demand of variable-frequency equalization is more complex. A more thorough study of the compactness of the neural model for variable equalization frequency, difference transmission channels model is necessary.

The proposed LSTM equalizer is based on supervised learning which need a consistent paired training dataset for training the neural networks to learn the non-linear mapping. This need the procedure of data collection and measurement. Although, this is also necessary for FFE-DFE based channel equalization. As recent advance of the semi-supervised/unsupervised learning raising\cite{Zhai_2019_ICCV,9086055}, there are potential chance for applying those self-adaptive learning approaches for signal equalizer design which may lighten the work of training dataset construction. 

\section{Conclusion}
\label{sec:conclusion}
In this paper, we have proposed a novel recurrent feedback equalizer based on LSTM neural network. The LSTM network is trained offline with a collected signal sequence dataset. We have shown that our approach outperforms than commonly used FFE-DFE signal equalization methods with meaningful metrics. The proposed LSTM equalizer is adaptable for variant-frequency signal equalization with augmented deep learning structure. Furthermore, the LSTM equalizer is amenable for hardware/software implementation with flexible portability for different channels signals equalization.

Future research will also address weakly-supervised and unsupervised learning approaches for next generation AI based equalizer design. Moreover, the compatibility of the neural learning based approaches towards the high-frequency, low-cost, low-power and low-latency signal equalization is also needs to be studied.

\bibliographystyle{ieeetr}
\bibliography{ref.bib}
\end{document}